\documentclass{PoS}

\usepackage{amsmath}
\usepackage{amsfonts}
\usepackage{amssymb}
\usepackage{latexsym}
\usepackage{graphicx}
\usepackage{float}
\usepackage{wrapfig}
\usepackage{epsfig}

\newcommand{\be}{\begin{eqnarray}}
\newcommand{\ee}{\end{eqnarray}}

\newcommand{\nn}{\nonumber }

\newcommand{\bmat}{\left ( \begin{array}{cc} 
}
\newcommand{\emat}{\end{array} \right )}

\newcommand{\Tr}{\rm Tr}

\newcommand{\eins}{\leavevmode\hbox{\small1\kern-3.8pt\normalsize1}}
\newcommand{\vect}{\left ( \begin{array}{c}}
\newcommand{\evect}{ \end{array} \right )}
\newcommand{\bmini}{\begin{minipage}} 
\newcommand{\emini}{\end{minipage}} 

\title{Chiral Symmetry breaking in Bosonic Partition Functions}

\ShortTitle{Bosonic Partition Functions}

 \author{M. Kellerstein\\
        Stony Brook University\\
        E-mail: \email{moshe.kellerstein@stonybrook.edu}}

\author{K. Splittorff\\
        Niels Bohr Institute\\
        E-mail: \email{split@nbi.dk}}

\author{\speaker{J.J.M. Verbaarschot}\\
        Stony Brook University\\
        E-mail: \email{jacobus.verbaarschot@stonybrook.edu}}

\abstract{
The behavior of quenched Dirac spectra of two-dimensional lattice QCD is 
consistent with spontaneous chiral symmetry breaking which is forbidden 
according to the Coleman-Mermin-Wagner theorem. One possible resolution of this
paradox is that, because of the bosonic determinant  in the
partially quenched partition function, the conditions of this theorem 
are violated allowing for spontaneous symmetry breaking in two dimensions
or less.  This goes back to  work by Niedermaier and Seiler on nonamenable 
symmetries of the hyperbolic spin chain  and earlier work by two of
the auhtors on bosonic 
partition functions at nonzero chemical potential.   
In this talk we discuss chiral symmetry breaking for
the bosonic partition function of QCD at nonzero isospin chemical potential and
a bosonic random matrix theory at imaginary chemical potential and compare
the results with the fermionic counterpart.
In both cases  the chiral symmetry group of the bosonic partition function
is noncompact. }

\FullConference{The 33rd International Symposium on Lattice Field Theory\\
		14 -18 July 2015\\
		Kobe International Conference Center, Kobe, Japan*}

\begin{document}

\section{Introduction}
According to the celebrated Coleman-Mermin-Wagner theorem, continuous
symmetries cannot be broken spontaneously in two or less dimensions. In 
essence, the reason is that because of the fluctuations, the order parameter
averages to zero. However this theorem does not apply to non-compact 
symmetries. This was pointed out by Niedermaier and Seiler \cite{niedermaier}, who argued that nonamenable symmetries are necessarily broken
spontaneously in two dimensions or less. Nonamenable Lie groups are Lie groups
for which no invariant mean exists such are for example noncompact semi-simple
Lie groups -- it cannot exist because of the 
divergent group volume.  As emphasized in particular by Seiler \cite{seiler}
spontaneous symmetry breaking of noncompact symmetries
is unavoidable 
in any dimensions of space-time.

Noncompact symmetries are an essential ingredient of the spectral analysis of
disordered systems. The reason is that the resolvent is given by the derivative of
a ratio of determinants, e.g. in QCD
\be
G(z) = \Tr \left \langle \frac 1{D+z} \right \rangle= \left . \frac d{dz'}\right |_{z'=z}
\left \langle \frac {\det(D+z')}{\det(D+z)}{\det}^{N_f}(D+m)\right\rangle,
\ee
where $D$ is the anti-Hermitian QCD Dirac operator and $N_f$ is the number of
flavors with quark mass $m$.
Let us consider the simplest case, which is the quenched limit ($N_f =0$).
Then the bosonic determinant can be represented as (for ${\rm Re}(z) > 0$)
\be
\frac 1{\det(D+z)}=\int d\phi_1 d\phi_2d\phi_1^* d\phi_2^*  
\exp \left [- \vect \phi_1^* \\ \phi_2^*\evect^T 
\bmat z & id \\id^\dagger & z \emat \vect \phi_1 \\ \phi_2\evect \right ].
\ee
The axial symmetry is given by 
\be
 \vect \phi_1 \\ \phi_2\evect \to \bmat e^s & 0 \\ 0& e^{-s}\emat \vect \phi_1 \\ \phi_2\evect \qquad {\rm and } \qquad
 \vect \phi_1^* \\ \phi_2^*\evect \to \bmat e^s & 0 \\ 0& e^{-s}\emat \vect \phi_1^* \\ \phi_2^*\evect
\ee
with $s$ real. Note that the U(1) symmetry of the action is not a symmetry
of the partition function because
such transformations will violate the complex conjugation property of the $\phi_k$ which is required
to have convergent integrals. Rather we observe that the axial symmetry group
is noncompact and given by Gl(1)/U(1) \cite{Sener-1,Osborn-1,Damgaard-1}. This argument can be easily extended
to include $N_f$ flavors. In the case of the fermionic partition function, which 
can be represented as the average of a Grassmann integral,
convergence is not an issue and
the axial symmetry group can be chosen to be the compact U(1) group. The noncompact
transformation is a symmetry group as well but would lead to divergent
contributions if pushed forward onto the bosonized degrees of freedom
(see \cite{sener2}).

 In \cite{mario-2d} we  analyzed  Dirac spectra of the naive QCD Dirac
operator in two dimensions, and found the same degree of agreement with 
random matrix predictions as was the case for QCD in four dimensions
\cite{mario-2d}. One possible explanation could be that states are localized
but that the localization length is much larger than the size of the box.
However,  the volumes considered 
\cite{mario-2d} were too small to confirm this possibility.

Alternatively, according to
the arguments of Niedermaier and Seiler, the axial symmetry group, because
of its noncompactness, is always broken spontaneously also in dimensions of two
or less, and it is possible to have  extended  quantum states.
 However, the conditions for  this statement to be valid
remain unclear. For example,
chiral symmetry may be restored in the presence of an external field such
as an (imaginary) chemical potential, and this is also expected to
occur for a bosonic
partition  function.  We also note that it is possible to have a nonzero
density of states near zero without spontaneous symmetry breaking
but rather with localized wave-functions 
\cite{mckane-stone}. What happens depends on the renormalization group 
flow but this is not within the scope of this work.

In this paper we address a simpler question: we explore the  difference
between fermionic and bosonic partition functions. We
consider QCD at nonzero isospin chemical potential as well as
 a random matrix model at nonzero (imaginary) chemical potential
\cite{andy,lehner}, and analyze the phase diagram of the corresponding fermionic
and bosonic partition function.

\begin{figure}[b!]
\centerline{\includegraphics[width=7cm]{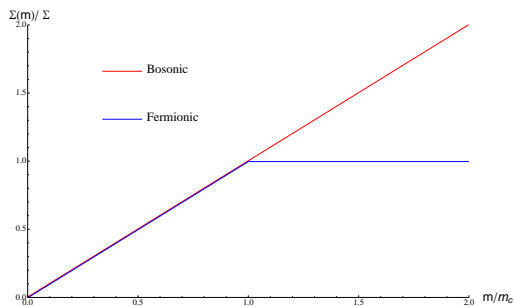}}
\caption{The mean field result for the mass dependence of the normalized chiral condensate for the
phase quenched bosonic (red) and fermionic (blue) QCD  partition function
at nonzero chemical potential $\mu$ 
versus the quark mass in units of the critical quark mass, $m_c$, for
which a condensation transition occurs. \label{fig1}}
\end{figure}

\section{Phase Quenched QCD}
Some time ago we analyzed \cite{split-bos} the difference between the fermionic and 
bosonic partition function in the $\epsilon$-domain of phase quenched QCD at nonzero chemical potential.
The fermionic partition function  given by 
\be
\left \langle \det (D+m+i\mu\gamma_0) \det (D+m-i\mu\gamma_0)\right \rangle.
\ee
This partition function is well understood \cite{SS}.
It is QCD at nonzero isospin
chemical potential with pion condensation for $\mu>m_\pi/2$.
In the $\epsilon$-domain of the QCD  partition 
given by an integral over  SU(2)
\be
Z = \int_{U\in {\rm SU(2)}}\exp[-\frac 14 V \mu^2 F_\pi^2 \Tr [U,\tau_3][U^{-1},\tau_3]
 + \frac 12 V \Sigma \Tr M(U+U^{-1})].
\ee
The SU(2) flavor symmetry of the fermionic partition function is broken to
U(1) by the chemical potential, and the residual $U(1)$ symmetry is
broken spontaneously by the formation of a pion condensate.
In the normal phase for $\mu < m_\pi/2$, the saddle point of the chiral Lagrangian
is at $U = 1$ while for  $\mu > m_\pi/2$ we are in a condensed phase with one
exactly massless Goldstone boson \cite{SS}.
The partition function is $\mu $-independent for
$\mu < m_\pi/2 $ with a mass independent chiral condensate while
it increases linearly in $m$ for 
$m_\pi/2< \mu$ (see Fig. \ref{fig1} for mean field results). 

The corresponding bosonic partition function is given by
\be
\left \langle \frac 1{\det (D+m+i\mu\gamma_0) \det (D+m-i\mu\gamma_0)}\right \rangle.
\ee
When $\lambda$ is an eigenvalue of $D+i\mu\gamma_0$, then $\lambda^*$ is
an eigenvalue of $D-i\mu\gamma_0$.  Expressing the average as an integral over
the spectral density, this
results in a logarithmic divergence
for  $\lambda \to m$
\be
\int_{C_\epsilon(m)} \frac{d\lambda d\lambda^*}{(\lambda-m)(\lambda^*-m)}\sim \log \epsilon,
\ee
where $C_\epsilon(m)$ is the region between a unit circle centered at $m$ 
and a circle of radius $\epsilon$ also centered at $m$. This divergence
of the partition function
can be regularized as
\be
\left \langle {\det}^{-1} \bmat D+m+i\mu\gamma_0 &\epsilon \\
\epsilon  & D+m-i\mu\gamma_0 \emat \right \rangle.
\ee
Using this regularization, the static part of the bosonic phase quenched partition function
takes the form
\cite{split-fac}
\be
Z = \int_{Q\in {\rm Gl(2)/U(2)}}\frac{dQ}{{\det}^2 Q}\theta(Q) \exp[-\frac 14 VF_\pi^2 \mu^2  \Tr[Q,\tau_3][Q^{-1},\tau_3] + \frac i2
V\Sigma \Tr  M (Q -I Q^{-1}I )],
\ee
with
$
M=\epsilon + m\tau_1$  and $I = -i\tau_2$.
After  the transformation $Q = i \tilde Q\tau_1$ the partition function
can be written as
\be
Z = \int_{-i\tau_2 \tilde Q\in {\rm Gl(2)/U(2)}}\frac{d\tilde Q}{{\det}^2 \tilde Q}
\theta(i\tilde Q\tau_1) \exp[-\frac 14V F_\pi^2\mu^2   \Tr[\tilde Q,\tau_3][\tilde Q^{-1},\tau_3]  - \frac 12
V \Sigma \Tr (M_\epsilon \tilde Q + M_{-\epsilon} \tilde Q^{-1} )
\ee
with
$
M_\epsilon =M \tau_1. 
$
Apart from the measure, for $\epsilon \to 0$ the bosonic and fermionic chiral Lagrangian
have the same functional dependence. At nonzero $\mu$ the U(2) symmetry of
the bosonic partition function is also broken to U(1). This
residual $U(1)$ symmetry is
broken spontaneously by the condensate resulting in a massless mode.
For the bosonic theory this is the case for any value
of $\mu$ while for the fermionic theory the U(1) symmetry  is restored
for  $\mu < m_\pi/2$.

The phase transition of the fermionic partition function occurs
at the point where the saddle point hits the boundary of the manifold. In the
bosonic case there is no boundary and the normal phase does not occur. 
Therefore, for any $\mu > 0$ we have a massless charged boson with nonzero
isospin charge
that condenses.
In Fig. \ref{fig1} we show the mass dependence of the chiral condensate for both
partition functions. In the fermionic case, there is a phase transition at
$m = m_c$ which is the quark mass  for which $\mu =m_\pi/2$,  while in the bosonic
case there is no such transition.

In terms of $\tilde Q$ the charged states reside in its off-diagonal
matrix elements, which are the diagonal matrix elements of $Q$. Indeed 
they contain a massless mode parameterized by
$ Q \to \exp[\tau_3 \alpha] Q \exp[\tau_3 \alpha]$ in the parameterization of
\cite{split-fac}.

What we conclude from this example is that spontaneous symmetry breaking
can persist in the bosonic theory while it is restored in the
fermionic theory for the same value of the parameters.

\section{Bosonic versus Fermionic one-flavor Partition Functions}
In this section we consider the one-flavor QCD partition function. Before analyzing
the random matrix model at nonzero imaginary chemical potential, we first remind the reader
of the $\epsilon$-limit of the 
bosonic and fermionic partition functions of one-flavor QCD at $\mu =0$.

\subsection{One Flavor Partition Function at Zero Chemical Potential}

At fixed topological charge $\nu$, the $\epsilon$-limit of the QCD partition function
 has a residual U(1) covariance so that
it is given by \cite{GL,LS}
\be
Z^{N_f =1}_\nu(m) &=& \int_{U\in {\rm U(1)}} dU {\det}^{\nu} U e^{\frac 12 mV\Sigma {\rm Tr}(U+U^{-1})}
=\frac 1{2\pi}\int_{-\pi}^\pi d\theta e^{i\nu\theta} e^{mV \Sigma\cos \theta}
= I_\nu(mV\Sigma).
\ee
The bosonic partition function is obtained by replacing the U(1) integral
by a Gl(1)/U(1) integral and is thus given by
\be
Z^{N_f =-1}_\nu = \int_{U\in {\rm Gl(1)/U(1)}} dU {\det}^\nu U e^{-mV \Sigma{\rm Tr}(U+U^{-1})}
= \int_{-\infty}^\infty ds e^{\nu s} e^{-mV\Sigma \cosh s} 
= K_\nu(mV\Sigma).
\ee
In both cases, for $\nu \ne 0$ the chiral condensate diverges  as $|\nu|/m$ for $m \to 0$.
For $\nu = 0$, the fermionic condensate vanishes in the chiral limit while the bosonic
condensate diverges as $1/m$. At fixed $\theta$ angle the condensates in each
sector are weighted by $Z_\nu/Z$ so that in the chiral limit only the
contributions for $\nu = \pm 1$ are nonvanishing in the fermionic case. In the
bosonic case, the sum over $\nu$ is divergent. Presently, it is not clear if the bosonic
partition function can be defined at fixed $\theta$.

\begin{figure}[b!]
\centerline{\includegraphics[width=6cm]{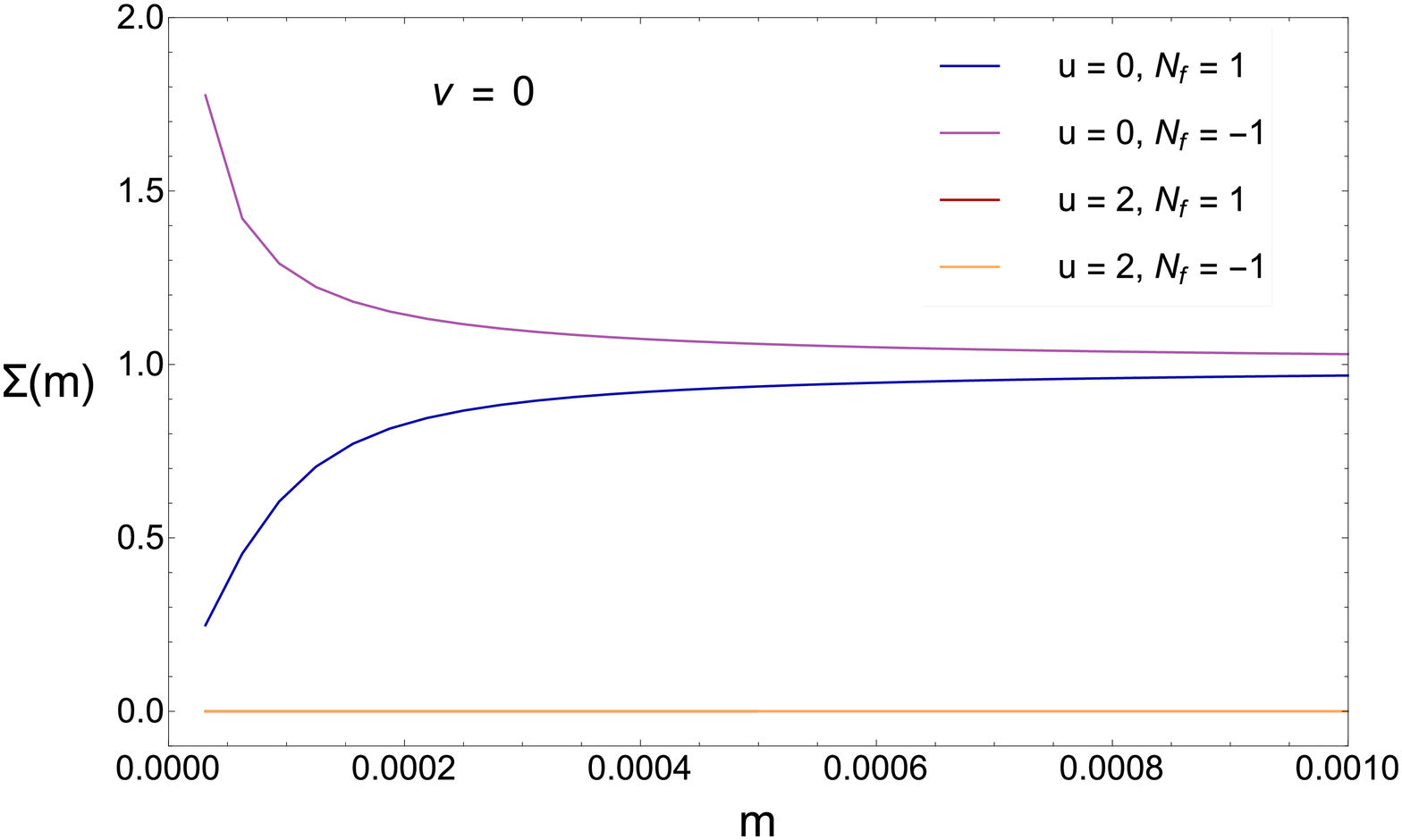}\hspace*{1cm}
\includegraphics[width=6cm]{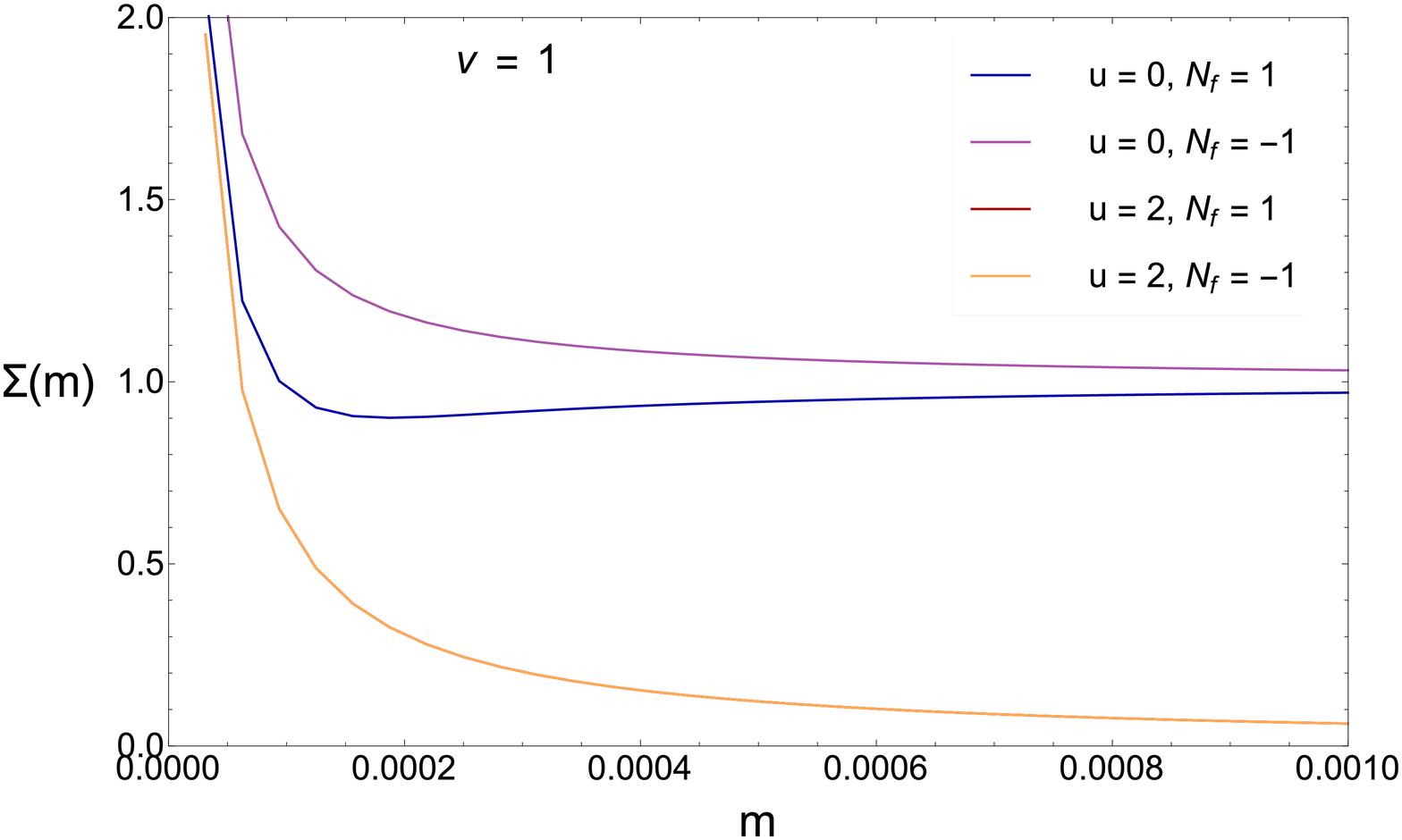}}
\caption{The mass dependence of the chiral condensate for the bosonic and fermionic partition function for $\nu=0$ (left) and $\nu=1$ (right). We show results for
  $u =0$ and $ u=2$.}
\label{fig2}
\end{figure}
\subsection{Chiral Random Matrix Theory at Imaginary Chemical Potential}

In this subsection we study a chiral random matrix model at imaginary 
chemical potential. The Dirac operator is given by \cite{andy,lehner}
\be
D = \bmat z & id +iu \\id^\dagger +iu & z \emat .
\ee
The partition function is given by the expectation
value of the determinant of $D$ averaged over Gaussian distributed matrix
elements $d$. In general,
  $d$ is a $ n\times (n+\nu)$ matrix so that it  has $ \nu$ 
zero modes. In the thermodynamic limit
we keep $N\equiv 2n+\nu$ fixed.
This model was first studied in \cite{andy} where
it was shown that it has a second order chiral phase transition with 
mean field critical indices. For imaginary $u$, or real chemical potential, 
the partition function
corresponding to $D$ is a model for QCD at nonzero baryon chemical potential
with a phase transition to a phase with a nonzero baryon density
\cite{stephanov}. If $u$ is interpreted as the lowest Matsubara frequency, it is also
a mean field model for the chiral phase transition as a function of
temperature \cite{andy}. 
The normalization can be chosen such that 
the chiral condensate is normalized to one for $u=0$ and becomes
zero at $u=1$.
 For imaginary $u$, using the same units, the phase transition is at
$u= 0.52$ though \cite{stephanov}.

 The fermionic partition function was worked out in \cite{andy,adam-mu,lehner}.
 For one-flavor the expression for finite
$n$ can be simplified to
\be
Z_\nu^{N_f=1}(m,u)= \int_0^\infty ds s^{\nu+1}  I_\nu(2n m s\Sigma) (s^2+u^2)^{n} e^{-n(s^2+m^2)}.\nn
\label{zfer}
\ee
 For $u<1$, the chiral condensate is given by $\Sigma(u) =\sqrt{1-u^2}$
 while for $u>1$ the condensate vanishes in the chiral limit
 (both for $\nu =0$). For small nonzero
$m$ it is given by $\Sigma(u) \sim mu/\sqrt{u^2 -1}$.

The bosonic partition function can be reduced to a one dimensional integral
\cite{keller}
\be
Z^{N_f=-1}_\nu(m,u) &=&                      
m^{-\nu}\int_0^\infty \frac{ds}{s^{\nu+1}} e^{-sn m^2/2}
\frac{1}{(1/s+1/\Sigma^2)^{n+\nu}} e^{- n u^2/(1/s+1/\Sigma^2)}
.
\label{zbos}
\ee
This partition function also shows a phase transition to a chirally restored phase
at $u=1$. For $u<1$ we find the same expression for the chiral condensate as
in the fermionic case but for $u>1$ we obtain a slightly different result,
$\Sigma(u) \sim m/\sqrt{u^2 -1}$ (also for $\nu =0$).  The main difference between the fermionic
and the bosonic partition function is that for $\nu= 0$ the
chiral condensate in the first case  vanishes for
$ m\to 0$ at fixed $n$, while in the case of the  bosonic partition
function it  diverges as $1/m$ for $m\to 0$ at fixed $n$. In this sense the
chiral symmetry of the bosonic partition function can also  broken in the chiral
limit. 

\begin{figure}[t!]
\centerline{\includegraphics[width=6cm]{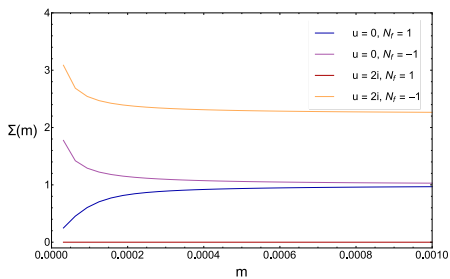}}
\caption{The mass dependence of the chiral condensate of real and imaginary chemical potential
(see legend). The chiral condensate of the bosonic partition function remains nonzero
also at large values of the chemical potential (orange curve).} \label{fig3}
\end{figure}

A second difference between the bosonic and fermionic partition function concerns
its
properties under analytical continuation. Replacing $u\to i\mu$ in Eq. (\ref{zfer})
reproduces the result for the random matrix model at nonzero real chemical potential
\cite{stephanov,adam-mu}.
The bosonic partition function can also be continued naively to imaginary $u$. In Fig. \ref{fig3}
we show  that in this case the chiral condensate remains finite
 for parameter values for which the chiral condensate of the fermionic partition
function vanishes in the thermodynamic limit. Whether
the analytical continuation in $u$ is valid for the bosonic partition function as well
remains to be determined
\cite{keller}.

\section{Conclusions}
We have compared spontaneous chiral symmetry breaking for bosonic and fermionic
partition functions, and have studied phase quenched QCD and a
one-flavor random matrix theory at
imaginary chemical potential.
In the first case,  a phase transition of the fermionic partition function
occurs when the saddle point reaches
the boundary of the manifold. This does not happen for
bosonic partition function when the saddle point manifold is noncompact,
and the residual U(1) symmetry is always broken spontaneously.
In the second example,  an axial symmetry restoration phase transition
occurs when the minimum of the effective potential trivializes.
The two partition functions have the same phase diagram but
may behave differently under analytical continuation.

\vspace*{0.2cm}
\noindent {\bf Acknowledgements.}
This work was supported by U.S. DOE Grant No. DE-FG-88ER40388 (MK and JJMV)
and the Sapere Aude program of The Danish Council for Independent Research (KS).

\end{document}